\documentstyle[11pt,newpasp,twoside,epsf]{article}
\newcommand{\HI}{\protect\normalsize H\thinspace\protect\footnotesize I\protect\normalsize}

\reversemarginpar

\begin{document}

\title{IR-TF Relation in the Zone of Avoidance with 2MASS}
\author{Nicolas Bouch\'e, Stephen Schneider }
\affil{University of Massachusetts, Department of Astronomy, Lederle
Graduate Tower, Amherst, MA 01003, USA}

\begin{abstract}
Using the Tully-Fisher (TF) relation, one can map the peculiar velocity field
and estimate the mass in regions such as the Great Attractor. 
The Two Micron All Sky Survey (2MASS)
is surveying the full sky in J, H and K bands and
has the great advantage that it allows us to detect galaxies more
uniformly to lower Galactic latitude than optical surveys.
This study shows the feasibility of using
2MASS and the TF relation to probe large-scale structure.
We found  that
 (i) we can use axial ratio $b/a$ up to 0.5;
 (ii) intrinsic extinction is present (up to 0.5 mag at J, 0.1 mag at K);
(iii) the zero-point of the TF relation is independent of the 2MASS magnitude measurement;
(iv) the 2MASS K-band 20th mag/arcsec$^2$ isophotal aperture magnitude produces  the best TF relation;
and (v) there is no type dependence of the residuals. 
%(vi) We will also present our results at low Galactic latitude.
\end{abstract}

\section{Introduction. The Tully-Fisher Relation}
In 1977, Tully and  Fisher discovered
that there is a good correlation between maximum rotational velocity, 
usually given by the global neutral hydrogen 
line profile width $\Delta V$\ (corrected for inclination)
and the absolute magnitude ${\rm M}$ for late-type galaxies
(Tully \& Fisher 1977).  

Observationally, the correlation (luminosity vs.~maximum velocity) 
gives $L \propto \Delta V^{\alpha}$ 
with $\alpha$ in the range of 2.5-3 at optical wavelengths.
There is a general trend that at longer wavelengths, the slope steepens.
At infrared wavelengths, the slope ($-2.5\times\alpha$ in magnitudes)
climbs to about  $-10$ (or $\alpha \sim 4$)
which is close enough to the Faber-Jackson relation
$L \propto \sigma_v^4$
to suspect a possible common physical origin.
\par
The origin of the TF relation lies (in the very roughest terms)
in the virial theorem. Assuming a constant mass-to-light ratio ($M/L$) and
constant surface brightness (I), one can rewrite the virial theorem as
\begin{equation}
L \propto V^4 [(M/L)^2I]^{-1} 
\end{equation}
The zero-point is directly related to basic physical properties as
can be seen if  rewritten as $-2.5 \times \log[I(M/L)^2]$ or 
$\log [(M/L)\Sigma]$. $\Sigma$ is the surface mass density which
is a strong function of the angular momentum ($\Sigma \propto M^7/J^4$) whereas
the mass-to-light ratio gives information on the stellar content.
This  can be used to put strong constraints on galaxy formation theory 
(see Steinmetz \& Navarro 1999; Koda et al.~2000).
	
Zwaan et al.~1995 also have shown that low surface brightness  
late-type galaxies follow the same B-band TF relation as the high surface
brightness galaxies, even though
their mass-to-light ratios are very different as well as the ratio 
of dark to luminous matter (de Blok \& McGaugh 1997), but this is
still somewhat controversial (see O'Neil et al.~2000).

The HST Key Project group on the extragalactic distance scale has used the
TF relation to infer the Hubble constant (${\rm H}_0$) 
(see Sakai et al.~2000 for a
recent discussion).
Another application is to use the TF relation to measure the peculiar velocity field of galaxies.
This provides an estimate of the mass density on large scales and of the density
of the universe through $\Omega^{0.6}/b$.
	
Using near-IR for TF studies has two advantages: (1) it is  sensitive to old red stars
and less sensitive to young blue stars.
In other words, it is fairly independent of current star formation,
so it  provides a measure of current
stellar content that is, presumably, more closely tied to the total mass
of the galaxy and hence its kinematics;
(2) uncertainties due to interstellar absorption in both our own Galaxy
and in the observed galaxy are greatly reduced.

This project shows the feasibility of using 2MASS (Huchra et al.~these
proceedings; Jarrett et al.~2000)
 and \HI\ data to study large scale
structure. We studied  the Pisces-Perseus (PP) supercluster region	
and the Great Attractor (GA) region in the Zone of Avoidance. In particular, we ask the following questions regarding 2MASS: \\
\indent (i) Which band (J, H or K) generates the best TF relation?\\
\indent (ii) What magnitude measurement gives the tightest TF relation?\\
\indent (iii) Can we use 2MASS data at low Galactic latitude?\\
and regarding the TF relation:\\
\indent (iv) What is the highest axial ratio $b/a$ that we can use?\\
\indent (v) Is there any internal extinction?\\

The remainder of this paper is organized as follows. 
In section~2, we present our data sets. The method is described in section~3.
Our results are presented in section~4 and summarized in section~5.

\section{Data Sets}
We construct two data sets. The first (PP) data
set consists of \HI\ data selected from the
Huchtmeier \& Richter catalog (Huchtmeier \& Richter 1989).
Because we found substantial scatter was introduced when using line widths measured
in different ways, we select only papers with Giovanelli and collaborators.
The data was also restricted  to the Arecibo telescope and measurements made at
50\% of the peak. The sample contains $\sim 2700$ objects.
We cross-matched 2MASS data with the Giovanelli subsample. There were 502 matches
as of spring 1999~\footnote{As of January 2000, there are about 3 times as many
matches, but the analysis is not completed.}.
This sample  covers the sky from $0^{\rm h}$ to $5^{\rm h}$ in right ascension 
and $15\deg$ to $35\deg$
in declination which corresponds to the Pisces-Perseus supercluster region.

We also used \HI\ data from Giovanelli et al.~1997
from their extensive survey of the I-band TF in clusters.
We again cross-matched this subsample with the
2MASS database and obtained 300 matches from 22 different clusters (hereafter {\it cluster} sample).
We took into account the  different offsets from the Hubble flow (column (4) in 
Table 2 of Giovanelli et al.~1997) for each cluster. This sample also
contains inclination information both from  I-band photometry and from 2MASS which
enabled us to perform checks on the 2MASS estimate of the axial ratio.

\section{Method and Error Budget}

\subsection{Sub-Sample}
In most cases, we used a subsample of galaxies chosen to have the following
properties:
(i) axial ratio $b/a$ less than 0.5 (the axial ratio used was the 
``super coadd'' axial
ratio, keyword {\it sup\_ba} in the 2MASS database,
which is determined from the coadded J, H and K images);
(ii) \HI\ flux minimum of 1 Jy km s$^{-1}$ in order to remove low SNR \HI\ data;
(iii) magnitude cutoff of K=13 mag (in the PP sample, we had a cutoff 
on the magnitude errors ($\sigma=0.1$) which corresponds to a cutoff in 
magnitude of $\sim 13$ determined from the dependence of the magnitude 
errors with magnitude);
(iv) excluding S0's and ellipticals (in general, the $b/a$
constraint and \HI\ detection are sufficient to ensure that no ellipticals 
are in the subsample), and
(v) redshifts between $z_{min}$ and $z_{max}$ to examine galaxies in 
different  distance ranges.

Figure~1 shows the TF relation for the {\it cluster} sample. 

\subsection{Corrections}
In order to improve the quality of the TF relation, a number of corrections are 
usually made  for turbulent motion, instrumental broadening, and extinction. 
These corrections have a number of adjustable parameters that we investigate here to
determine which values or methods work best with the 2MASS data.

We have corrected for (i) instrumental broadening following Bottinelli et al.~(1990) and
Giovanelli et al.~(1997); (ii) turbulent motion according to the scheme
proposed by  Tully \& Fouqu\'e 1985 (this correction is small and
did not improve the scatter of the TF relation);
(iii) inclination, i.e. \mbox{$v_{rot}=\frac{W_R}{2 \sin i}$} where $W_R$ is the observed
velocity width $W_{50}$ corrected for instrumental broadening and $i$
is the disk inclination based on the axial ratio $b/a$ from 2MASS, and (iv)
photometric corrections such as the Galactic extinction (from COBE/ DIRBE maps,
Schlegel et al.~1998) and internal extinction\mbox{ $\Delta m=-\gamma \log b/a$} where
$\gamma$ characterizes the extinction for a highly inclined  
galaxy.  In general, $\gamma$ can be a function of galaxy luminosity or
velocity width (see Sakai et al.~2000 and references therein), 
but we treated it as a constant.

\subsection{Parameterization and Error Budget}

We parameterize the TF relation in the same way as Giovanelli et al.~(1997), $y={\rm a}+{\rm b}x$ where
$y$ is the absolute magnitude and $x$ is the logarithm of the rotation velocity, i.e.:
\begin{eqnarray}
x&=&\log \left [ \frac{W_R}{2 \sin i} \right ] - 2.5  \\
y&=&(m_{obs}-A-\Delta m -k_z -5 \log \frac{cz}{100{\rm h}_{100}} -25 ) 
- 5 \log {\rm h}_{100} \nonumber \\
 &=& M - 5 \log {\rm h}_{100}
\end{eqnarray}
\noindent
where h$_{100}$ is the Hubble constant in units of 100 km s$^{-1}$ Mpc$^{-1}$. 
The last term
$- 5 \log {\rm h}_{100}$ makes $y$ independent of the Hubble constant so it is easier 
to compare
results with other authors. $k_z$ is the $k$-correction and was generally ignored here
since the sample contains only nearby galaxies ($cz<10,000$ km s$^{-1}$). 

A standard $\chi^2$ fit to the data was performed using errors both along 
the $x$ and 
$y$ axes, $\sigma_x$ and $\sigma_y$ ({\it bivariate} fit).  $\sigma_x$ includes velocity width 
uncertainty as well as inclination errors, which dominate the overall error. $\sigma_y$
includes magnitude uncertainty from 2MASS, distance uncertainty, and an added
intrinsic scatter $\sigma_i$ which takes into account any ``cosmological'' scatter inherent in the
 TF relation. 
The $1\sigma$ uncertainty on the parameters (a and b) is given by the projection 
of the $1\sigma$ contour semi-major axis onto the parameter axis (Press et al.~1992). 

\begin{figure}
\plotfiddle{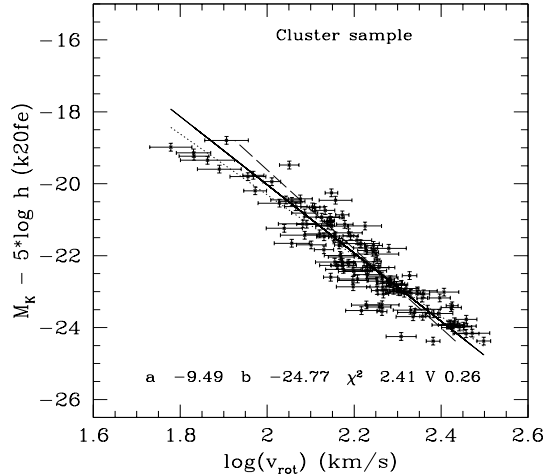}{6.0cm}{0}{36}{33}{-110}{-62}
\caption{TF relation for a sub-sample (130 objects) of the {\it cluster} data set 
with axial ratio $b/a  < 0.5$, \HI\ flux $>$ 1 Jy km s$^{-1}$, magnitude $m_K<13$.
The slope, a, and zero-point, b,
are shown as well as the  $\chi^2_{min}$ and the variance $V$. 
The dotted line is a
direct fit, the dashed line is an inverse fit and the bivariate fit 
 is the solid line with an intrinsic scatter $\sigma_i$ of 0.2 mag. No internal extinction
correction was made here.}
\end{figure}

\section{Results and Discussion}

2MASS produces many different magnitude measurements for each object (Jarrett et al.~2000). 
The different types of magnitude that we used are
(i) the 20th mag/arcsec$^2$ isophotal magnitude ({\it i20} in the 2MASS 
database~\footnote{
In order to find the proper keyword in the 2MASS database, one needs to add 
the prefix {\it j\_m\_, h\_m\_} or {\it k\_m\_} and the suffix 
{\it e} ({\it c}) for elliptical (circular) aperture.  });
(ii) the 21st mag/arcsec$^2$ isophotal magnitude ({\it i21});
(iii) the 20th mag/arcsec$^2$ K fiducial~\footnote{2MASS has two kinds of 
magnitude type, ``individual'' (i20 and j21 here) and ``fiducial'' 
(k20f, j21f and kf here) . 
``Fiducial'' means the aperture used for the photometry was selected at 
one band and applied to the other two. }
magnitude ({\it k20f});
(iv) the 21st mag/arcsec$^2$ J fiducial magnitude ({\it j21f}), and
(v) the Kron (Kron 1980) K fiducial ({\it kf}) magnitude which measures the 
flux within an aperture 2$r_1$ where $r_1$ is the first moment of the light 
distribution.
This type of  magnitude is thought to be less sensitive to observing
conditions and is thought to be a ``total'' measure of the integrated flux 
(Koo 1986).

First, which band J, H or K produces
the least scatter in the TF relation? Figure~2 shows the minimum value $\chi^2_{min}$
of the bivariate fit for each of the magnitude types. Figure~2
indicates that (1) the {\it 20th mag/arcsec$^2$ K fiducial magnitude} and Kron K fiducial produces the smallest scatter; (2)  in general K gives a tighter TF relation than J or H, 
considering that  H-band photometry of 2MASS is subject to uncertainty due to air glow; and
(3) elliptical apertures tend to give a better TF relation than circular ones,
although this was not consistently true for all datasets.

Since the different magnitude types have different measurement errors, the $\chi^2_{min}$
could look artificially low if the error estimates were inflated. However, we found no
effects of this sort.
Magnitude types with the largest
mean measurement error did not produce the smallest $\chi^2$.

\begin{figure}
\plotfiddle{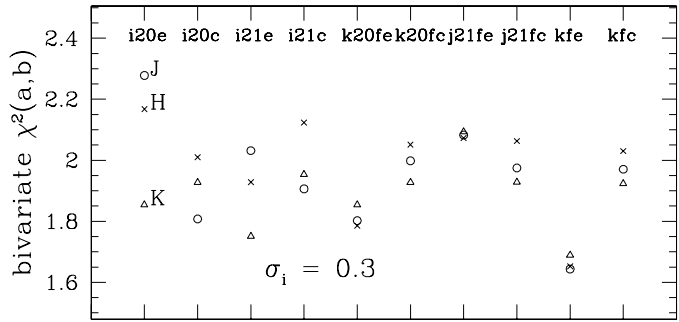 }{5.5cm}{0}{45}{45}{-140}{-120}

\caption{For each of the magnitude types, the reduced $\chi^2$ value of the fit is shown, where
the circles are for the J-band magnitude, the crosses are for the H-band, and the triangles are for the K-band.
Circular (elliptical) aperture magnitudes are indicated the suffix {\it c} ({\it e}).
This is for the {\it cluster} data set using 2MASS axial ratios for the inclination and
an intrinsic scatter $\sigma_i$ of 0.3 mag.
}
\end{figure}

Figure~3 shows the fitted values of the slope, a, and the zero-point, b, 
for each
of the magnitude types assuming an intrinsic scatter $\sigma_i$ of 0.2 for the {\it cluster}
sub-sample and using the 2MASS axial ratios for the inclination. 
First note that the
zero-point of the TF relation is well determined and does not depend on the magnitude used.
Second, the slope is more dependent on the magnitude type. However, the 
``fiducial''
magnitudes are well within 1-$\sigma$ of each other. 
Third, the slope tends to be 
steeper as the wavelength increases. This is often called the color-TF relation. 
The same conclusions can be reached using either the 2MASS or I-band 
inclinations
and different values of $\sigma_i$.

\begin{figure}
\plotfiddle{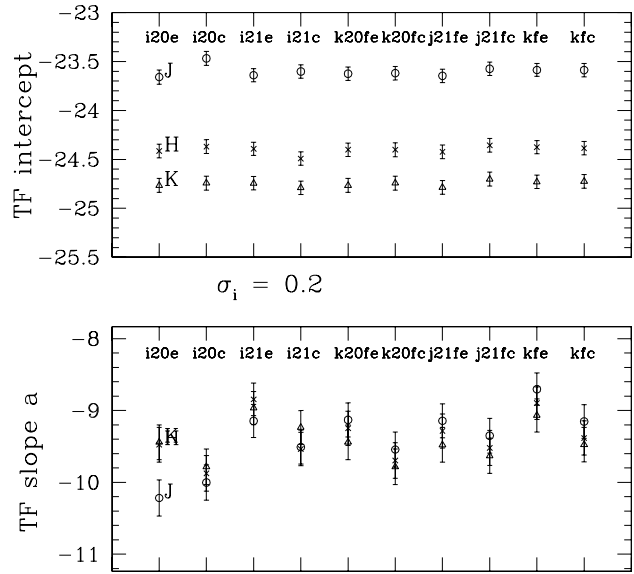 }{7cm}{0}{43}{43}{-130}{-89}

\caption{Top panel: Intercept of the TF fit for each of the magnitude types. 
For each  of the magnitudes, the circles are for the J-band magnitude, the crosses are for the H-band, and the triangles are for the K-band. 
Circular (elliptical) aperture magnitudes are indicated by the suffix 
{\it c} ({\it e}).
Notice that the zero points are all within 1-$\sigma$ of each other 
especially at K.
Bottom panel: Slopes of the TF relation for each of the magnitude types. 
Both panels are for
the {\it cluster} sub-sample using 2MASS axial ratios with no internal extinction correction.
}

\end{figure}

Figure~4 shows the value of the reduced $\chi^2$ as a function of $\gamma$ 
for the PP sample using the magnitude {\it k20fc} and an intrinsic scatter 
of 0.2 mag.
We find a minimum  for each band: $\sim 0.5$ mag at J, $\sim 0.3$ at H and 
$\sim 0.1$ at K,
although the amount of extinction at K is consistent with none. 
 These curves are consistent with our results from the {\it cluster} sample, however,
they are not fully reproduced when using the inclination from the I-band photometry.
Our results are consistent with a reddening of E(B-V) of 0.5 mag which gives a $A_\lambda$ of
0.45 mag, 0.28 and 0.18 for the three bands.

\begin{figure}
%\plotone{bouche4.eps}
\plotfiddle{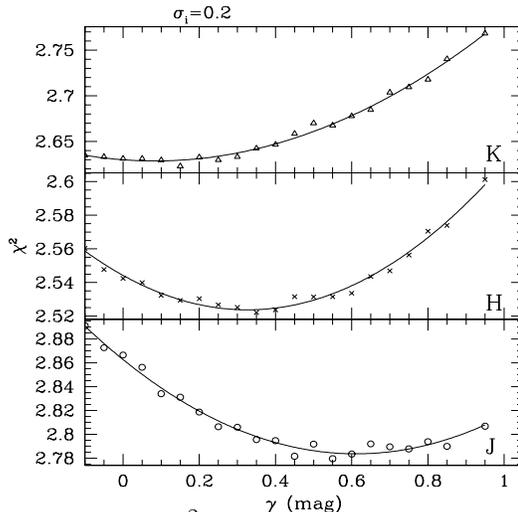}{6cm}{0}{37}{35}{-120}{-65}

\caption{Plot of the $\chi^2_{min}$  as a function of the parameter $\gamma$ for
the PP data set.
The panels correspond to J, H and K starting from the bottom. 
The open symbols are the data points while the solid line
is a fit of a second order polynomial. The amount of extinction
(at the minimum of each curve) 
decreases as the wavelength increases as one would expect.
\label{extinction}
}
\end{figure}

We did not see any correlation with type for the {\it cluster} data set while
the PP data set shows a slight correlation. A linear fit gives a non-zero 
slope ($-0.06$) with 1-$\sigma$ uncertainty of 0.03 for the PP data set,
such that both results are consistent with no type dependence (at the 95\% 
confidence level).

We have not tested for dependence of the TF relation on surface brightness 
because 
the current 2MASS galaxy processor is not very sensitive to low surface brightness disks. 
Attempts are underway to extract them in the future.

\section{Conclusions}

From our two samples (PP and {\it cluster} data sets), we found that
using K-band is generally preferable over the other bands. The K-fiducial 20th mag/arcsec$^2$
(and the K-fiducial Kron, although in the PP sample, the Kron K-fiducial 
does not produce
such a low $\chi^2$) seems to produce the least scatter. This is consistent with
the K-fiducial isophotal elliptical aperture magnitude being the most robust 
photometric measurement
(Jarrett et al.~2000). At low Galactic latitude, preliminary results suggests 
one would need
optical axial ratios because the high star density can mislead the 2MASS 
ellipse-fitting program.
The zero-point is well determined and is even independent of the magnitude 
types and of the sample.
The TF slope shows more scatter but it steepens with wavelength.
An interesting result is that we find internal extinction at each band, 
although the amount of extinction at K is consistent with none.
We find no type dependence in our samples but
we cannot yet conclude whether there is a surface brightness dependence.

Our results are summarized below for the PP sample (equations~4 \& 5) and
the {\it cluster} sample (equations~6 \& 7) 
using an intrinsic scatter of $\sigma_i=0.2$ and assuming $\gamma=0.1$ mag of internal extinction.
\begin{eqnarray}
	%for PP sample using plots.sm v1.3
M_K-5\log {\rm h}_{100}&=&( -9.90\pm 0.35)\cdot (\log v_{rot}-2.5) + (-24.70\pm0.10)  \\
        %for PP sample using plots.sm v1.3 
	% -9.635      0.3484      -24.76     0.08778       2.596    N=71
M_K-5\log {\rm h}_{100}&=&( -9.64 \pm 0.35)\cdot (\log v_{rot}-2.5) + ( -24.76 \pm  0.09)  \\ 
%\end{eqnarray}
%\begin{eqnarray}
	%For the {\it cluster} sample
	%using plots.sm v1.18
M_K-5\log {\rm h}_{100}&=&( -9.63\pm 0.27)\cdot (\log v_{rot}-2.5) + (-24.74\pm0.08) \\
	%M_K-5\log h&=&(-9.58\pm 0.29)\cdot (\log v_{rot}-2.5) + (-24.86\pm0.10)   
	 %For the {\it cluster} sample 
        %using plots.sm v1.18
	%-9.583      0.2929      -24.86      0.1041        2.78      0.2422 N=116
M_K-5\log {\rm h}_{100}&=&( -9.58 \pm  0.29)\cdot (\log v_{rot}-2.5) + ( -24.86 \pm  0.10) 
\end{eqnarray}
Equations~4 \& 6 are for the {\it k20fc} magnitude, while equations~5 \& 7 
are for the {\it k20fe} magnitude.
Both samples give very similar results. 

As part of future work to study large scale flows in the ZOA, 
we have recently collected \HI\ data at Arecibo for low Galactic latitude galaxies. Of
169 sources at $|b|<10\deg$, 72 were detected; of 147 sources at 
$10\deg<|b|< 20\deg$, 75
were detected. 
The galaxies were identified using the 2MASS galaxy processor, which holds
promise for identifying a large sample of galaxies within the ZOA.

{\bf Acknowledgments}
N. Bouch\'e is the recipient of a Five College Astronomy Graduate Fellowship
supported by the Mary Dailey Irvine Fund which allowed him to attend the meeting.


\begin{references}
\reference Bottinelli, L., Gouguenheim, L., Fouqu\'e, P., \& Paturel, G.
1990, \aaps, 82, 391
\reference de Blok, W.J.G., \& McGaugh, S.S. 1997, \mnras, 290, 533
\reference Giovanelli, R., Haynes, M.P., Herter, T., \& Vogt, N. 1997,
\aj, 113, 53
\reference Huchtmeier, W.K., \& Richter, O.G. 1989, \aap, 210, 1 
\reference Jarrett, T.H., Chester, T., Cutri, R., Schneider, S.,
Skrutskie, M., \& Huchra, J.P. 2000, \aj, 119, 2498
\reference Koda, J., Sofue, Y., \& Wada, K. 2000, \apj, 531, L17
\reference Koo, D.C. 1986, \apj, 311, 651
\reference Kron, R.G. 1980, \apjs, 43, 305
\reference O'Neil, K., Bothun, G.D., \& Schombert, J. 2000, \aj, 119, 1360
\reference Press, W.H., Teukolsky, S.A., Vetterling, W.T., \& Flannery, B.
1992, Numerical Recipes in C, (Cambridge: Cambridge University Press)
\reference Sakai, S., Mould, J.R., Hughes, S.M.G., et al.~2000, \apj, 529, 698
\reference Schlegel, D.J., Finkbeiner, D.P., \& Davis, M. 1998, \apj, 500,
525
\reference Steinmetz, M., \& Navarro, J.F. 1999, \apj, 513, 555
\reference Tully, B., \& Fisher, R. 1977, \aap, 54, 661
\reference Tully, B., \& Fouqu\'e, P. 1985, \apjs, 58, 67
\reference Zwaan, M.A., van der Hulst, J.M., de Blok, W.J.G., \& McGaugh, S.S.
1995, \mnras, 273, L35
\end{references}
\end{document}